\documentclass[journal=jctcce,manuscript=article]{achemso}

\usepackage{chemformula} 
\usepackage[T1]{fontenc} 
\usepackage{amssymb}
\usepackage{caption}
\author{Pjotrs \v{Z}guns}
\email{pjotrs.zguns@cfi.lu.lv}
\fax{+371 67132778}

\affiliation[ISSPUL]
{Institute of Solid State Physics, University of Latvia, Kengaraga street 8, LV-1063, Riga, Latvia}
\alsoaffiliation[MIT]
{Department of Materials Science and Engineering, Massachusetts Institute of Technology, 77 Massachusetts Avenue, Cambridge, Massachusetts 02139, USA}

\author{Inga Pudza}

\author{Alexei Kuzmin}
\email{a.kuzmin@cfi.lu.lv}

\affiliation[ISSPUL]
{Institute of Solid State Physics, University of Latvia, Kengaraga street 8, LV-1063, Riga, Latvia}

\title[Benchmarking CHGNet uMLIP Against DFT and EXAFS]{Benchmarking CHGNet Universal Machine Learning Interatomic Potential Against DFT and EXAFS: Case of Layered \ch{WS2} and \ch{MoS2}}

\abbreviations{uMLIP, DFT, EXAFS, MD }

\begin{document}

\begin{tocentry}
\includegraphics[width=0.8\textwidth]{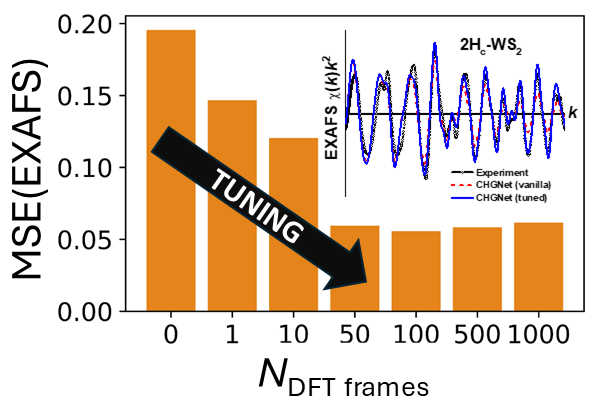}
\end{tocentry}

\begin{abstract}
Universal machine learning interatomic potentials (uMLIPs) deliver near \emph{ab initio} accuracy in energy and force calculations at low computational cost, making them invaluable for materials modeling. Although uMLIPs are pre-trained on vast \emph{ab initio} datasets, rigorous validation remains essential for their ongoing adoption. In this study, we use the CHGNet uMLIP to model thermal disorder in isostructural layered 2H$_c$-WS$_2$  and 2H$_c$-MoS$_2$, benchmarking it against \emph{ab initio} data and extended X-ray absorption fine structure (EXAFS) spectra, which capture thermal variations in bond lengths and angles. Fine-tuning CHGNet with compound-specific \emph{ab initio} (DFT) data mitigates the systematic softening (i.e., force underestimation) typical of uMLIPs and simultaneously improves alignment between molecular dynamics-derived and experimental EXAFS spectra. While  fine-tuning with a single DFT structure is viable, using $\sim$100 structures is recommended to accurately reproduce EXAFS spectra and achieve DFT-level accuracy. Benchmarking the CHGNet uMLIP against both DFT and experimental EXAFS data reinforces confidence in its performance and provides guidance for determining optimal fine-tuning dataset sizes.
\end{abstract}


\section{Introduction}

Machine Learning Interatomic Potentials (MLIPs)\cite{Behler2007NN, Bartok2010GAP,Thompson2015spectral,Shapeev2016tensor,Smith2017ANI, Wang2018deepmd,Drautz2019cluster} enable the computation of forces and energies with \emph{ab initio} accuracy but at a significantly lower computational cost. This advantage allows performing long and accurate Molecular Dynamics (MD) simulations to accelerate modeling of various material properties such as melting point\cite{Bartok2018Si,Jinnouchi2019onthefly}, phase transitions\cite{Jinnouchi2019MAPbI3,Fransson2023phase}, thermal transport\cite{Korotaev2019thermal,Verdi2021thermal}, ionic mobility\cite{Qi2021ionic,Zguns2024protons}, and more.

Universal MLIPs (uMLIPs), also known as foundational models, are particularly appealing\cite{choudhary2021ALIGNN,chen2022M3GNet,deng2023chgnet,merchant2023GNOME,batatia2023foundation,yang2024mattersim}. Pre-trained on large \emph{ab initio} datasets with up to ten million compounds, they demonstrate excellent performance and can be used as is, or fine-tuned with modest-sized \emph{ab initio} input\cite{deng2024overcoming} tailored for specific compounds. The circulation of uMLIPs started only recently, meaning their performance has not been extensively tested. Therefore, uMLIPs must be systematically compared to \emph{ab initio} calculations and experimental data\cite{Morrow2023validation,Yu2024assessment} to evaluate their performance and reliability.

In this paper, we rigorously test the accuracy of uMLIPs in capturing thermal disorder caused by lattice vibrations, evaluating uMLIP against the experimental Extended X-ray Absorption Fine Structure (EXAFS) spectra, which are sensitive to subtle variations of bond lengths and angles.\cite{Purans1999_Ge_thermal,Purans2008} EXAFS spectra result from the single- and multiple-scatterings of photoelectrons on the neighboring atoms, forming a complex fingerprint in X-ray absorption that reflects instantaneous atomic positions. MD simulations with uMLIPs, on the other hand, capture instantaneous positions influenced by thermal vibrations, enabling the computation of a theoretical EXAFS spectrum using multiple-scattering theory\cite{Rehr2000,Ankudinov1998}. By comparing the MD-derived spectrum with the experimental data, we can effectively test the accuracy of interatomic potentials\cite{Kuzmin2016,shapeev2022validation}.

We pose the following questions: How well does an original uMLIP capture structural features and thermal disorder? Does uMLIP performance improve significantly after being fine-tuned with minimal \emph{ab initio} input? What size dataset is required so that the fine-tuned uMLIP is indistinguishable from \emph{ab initio} force calculator? How does agreement with the experiment (EXAFS) change with the dataset size? By answering these questions, we aim to provide practical recommendations for uMLIP users, validated through both \emph{ab initio} calculations and EXAFS experiments.

We select the CHGNet package\cite{deng2023chgnet}, which has been trained on relaxation trajectories from the Materials Project database\cite{Materials_Project}. CHGNet is one of the best-performing uMLIPs\cite{matbench} among publicly available packages. Moreover, it not only computes forces and energies but also predicts magnetic moments of atoms, making it so far the only uMLIP that probes electronic structure features. Although we do not consider magnetic effects in this work, this feature makes CHGNet attractive for future studies.

\begin{figure*}[h]
	\centering
	\includegraphics[width=0.6\textwidth]{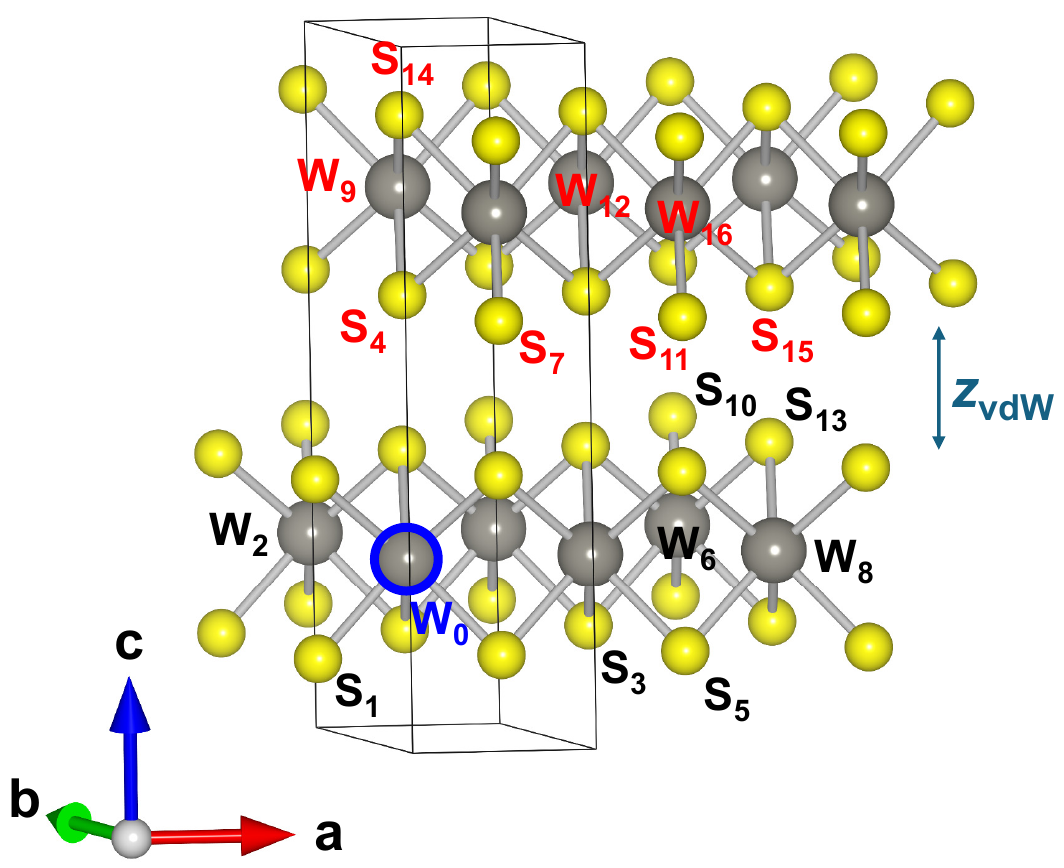}
	\caption{
    Crystallographic structure of hexagonal 2H$_c$-\ch{WS2}\cite{Schutte1987} (space group $P6_3/mmc$\cite{Schutte1987}), showing atoms located in the nearest 16 coordination shells around the W$_0$ atom (see also Table\ S2).
    The contribution of these atoms to the EXAFS spectrum is explicitly accounted for in this work.
    The vdW gap, $z_{\text{vdW}}$, is 3.019~\AA\ according to experimental data\protect\cite{Schutte1987}.  The lattice parameters\protect\cite{Schutte1987} are $a$ = 3.153~\AA\ and $c$ = 12.323~\AA,  with W–-S bond length $d_{\text{WS}}$ of 2.405~\AA.}
	\label{fig1}
\end{figure*}

We apply CHGNet to model thermal disorder in layered tungsten and molybdenum disulfides (2H$_c$-\ch{WS2}\cite{Schutte1987,Katzke2004polytypes} and isostructural
2H$_c$-\ch{MoS2}\cite{Schonfeld1983_MoS2}, Figure\ \ref{fig1}), high-symmetry compounds with interlayer van der Waals (vdW) coupling. By progressively fine-tuning CHGNet with an increasing number of \emph{ab initio} calculations, we assess how this affects the accuracy of energies, forces, and stresses. We then use the fine-tuned CHGNet to model thermal disorder through MD simulations, tracking how the agreement between the MD-derived EXAFS spectrum (W L$_3$-edge or Mo K-edge) and the experimental spectrum improves as the size of the fine-tuning dataset increases. This allows us to identify the minimum dataset size needed to achieve near \emph{ab initio} accuracy as well as accurately reproduce the experimental EXAFS spectrum.

\section{Methods}

\subsection{DFT calculations}
DFT calculations are performed within the Project Augmented Wave method\cite{PAW_method}, as implemented in the \texttt{VASP} code\cite{vasp_code_1,vasp_code_2,	vasp_code_3,vasp_code_4,vasp_code_PAW}. The Perdew--Burke--Enrzerhoff (PBE) exchange--correlation functional\cite{PBE_1,PBE_2} with D3 vdW correction\cite{D3_correction} is used. The kinetic energy cut-off is set to 520~eV, in consistency with Materials Project\cite{Materials_Project} settings. The following $\Gamma$-centered $k$-point grids ensure the total energy convergence within $\sim$0.05~meV/atom: $8 \times 8 \times 2$ for hexagonal unit cell of \ch{WS2} or \ch{MoS2} (6 atoms); $3 \times 3 \times 2$ for $3a \times 3a \times c$ supercell (54 atoms); $2 \times 2 \times 2$ for larger supercells (up to $6a \times 6a \times c$).

\subsection{Fine-tuning of pre-trained CHGNet potential}
Fine-tuning of pre-trained CHGNet potential (version 0.3.0\cite{deng2023chgnet}) is performed on short relaxation trajectories generated at seven isotropic strain states (from $\epsilon = -1.5\%$ to  $\epsilon = 1.5\%$, in $0.5\%$ increments), using $3a \times 3a \times c$ supercells (54 atoms). For each $\epsilon$,  100 relaxation trajectories are computed, initiated by randomly displacing atoms by 0.2~\AA\  and followed by 10 relaxation steps, resulting in a total of 7000 DFT frames (our tests show that fine-tuning on relaxation snapshots yields results comparable to those of fine-tuning on an equivalent number of MD snapshots). Fine-tuning on DFT-computed energies, forces, and stresses is performed using the Adam optimizer in CHGNet over five epochs with a batch size of 4, and an initial learning rate of $10^{-2}$. Most of the CHGNet layers are frozen, as tuning all layers offers no significant gain (see Table S3). The training, validation and test sets ratio is set to 8:1:1. The performance of the fine-tuned potential is further checked on extra test sets (Table S3).

Fine-tuning is performed on a server equipped with two 20-core Intel Xeon Gold 6148 processors and takes approximately 4000 seconds to complete for 7000 DFT frames in the tuning dataset.  Proportionally less time is required for smaller datasets, making the fine-tuning process relatively fast and unlikely to pose a computational bottleneck.

\subsection{Classical MD simulations}
Classical MD simulations are performed at 300~K in the NVT ensemble\cite{Abraham1986} using the Nosé--Hoover thermostat\cite{Hoover1985}, as available in CHGNet\cite{deng2023chgnet} via ASE implementation\cite{Larsen2017ASE}. Simulations use a timestep of 1~fs and an $8a \times 8a \times 2c$ supercell (768 atoms) with lattice parameters fixed to their respective relaxed values (Figure\ \ref{fig2}). The pressure in NVT simulations is about 0.5 GPa, with NPT ensemble simulations confirming that thermal expansion at 300 K compared to static structures is minor ($\Delta a/a \approx 0.2 ~\%$ and $\Delta c/c \approx 0.5 ~\%$), which is therefore neglected. Equilibration and production runs are at least 50~ps each, and 5000 MD snapshots from the production run are used to compute W L$_3$-edge or Mo K-edge EXAFS spectra. MD with the SWMBL-C force-field for \ch{WS2}\cite{BANDURA2023} is performed similarly, using GULP6.1.2 code\cite{Gale1997,GULP2003}, with lattice parameters $a = 3.161$\AA, $c = 12.331$\AA, a timestep of 0.5~fs, 20~ps equilibration and production runs, and 4000 MD snapshots for EXAFS spectrum.

\subsection{EXAFS calculations}
EXAFS calculations are performed on MD snapshots to obtain configuration-averaged spectra\cite{Kuzmin2009,KUZMIN2020rev}. We use \emph{ab initio} real-space FEFF8.5L code\cite{Rehr2000,Ankudinov1998}, incorporating multiple-scattering effects up to the fourth order\cite{Zabinsky1995,Ravel2005,Rehr2009} and scattering paths with a length up to 8~\AA, thus, including contributions from the sixteen coordination shells around the absorbing W (or Mo) atom (Figure\ \ref{fig1} and Table\ S2). The cluster potential is based on the average 2H$_c$-\ch{WS2} or 2H$_c$-\ch{MoS2} crystallographic structure within the muffin-tin approximation with 15\% overlap\cite{Rehr2000}, neglecting variations from atomic thermal motion. Inelastic effects are included using the energy-dependent Hedin--Lundqvist exchange-correlation potential\cite{Hedin1971}.

\subsection{EXAFS experiment}
X-ray absorption spectra at the W L$_3$-edge (10207~eV) of bulk 2H$_c$-WS$_2$ and at the Mo K-edge (20000~eV) of bulk 2H$_c$-MoS$_2$ are measured at 300~K at the DESY PETRA-III storage ring (6.08 GeV, 100 mA, top-up 480 bunch mode). Measurements are performed in transmission mode using undulator radiation from the P65 Applied X-ray Absorption Spectroscopy beamline\cite{Welter2019}, equipped with harmonic-reducing silicon mirrors, Si(111) monochromator, and ionization chambers for sample and reference foil detection. Tungsten(IV) sulfide (\ch{WS2}) and molybdenum(IV) sulfide (\ch{MoS2}) powders (Sigma-Aldrich, 99~\% and 98~\%, respectively) are confirmed by X-ray diffraction and Raman spectroscopy. Each powder is ground in an agate mortar, mixed with cellulose, and pressed into pellets using a hydraulic press, with a thickness optimized to yield the absorption edge jump of $\Delta \mu = 1$. The EXAFS spectrum, $\chi(k)k^2$, is extracted using the XAESA code\cite{XAESA}, following standard procedures\cite{Kuzmin2014}. The threshold energy $E_0$, used in the definition of the wavenumber $k = \sqrt{(2 m_\text{e}/\hbar^2) (E - E_0)}$ (where $E$\ is the X-ray photon energy, $m_\text{e}$ is the electron mass, and $\hbar$ is the reduced Planck's constant), is chosen to best align the experimental and theoretical EXAFS spectra\cite{Kuzmin2014}. The FTs are calculated using 10\% Gaussian function. Note, the FT peak positions differ from true crystallographic values due to the EXAFS phase shifts.

\section{Results and Discussion}

\subsection{Benchmarking uMLIP against DFT}

The original pre-trained CHGNet (hereafter referred to as vanilla CHGNet\cite{deng2023chgnet}) accurately predicts the W--S bond length, $d_{\text{WS}}$ = 2.421~\AA, and the in-plane lattice parameter, $a$ = 3.186~\AA\ (the experimental values\cite{Schutte1987} are 2.405~\AA\  and 3.153~\AA, respectively). However, it severely overestimates the thickness of the vdW gap, $z_{\text{vdW}}$ = 5.62~\AA\  (3.02~\AA\ in ref.\cite{Schutte1987}), leading to a significantly larger lattice constant $c = 17.54$~\AA\  (12.32~\AA\  in ref.\cite{Schutte1987}). This overestimation is expected, as vanilla CHGNet has been trained\cite{deng2023chgnet} on Density Functional Theory (DFT) data computed at the generalized gradient level\cite{PBE_1}, which does not account for vdW interactions.

\begin{figure*}[h]
	\centering
	\includegraphics[width=1.0\textwidth]{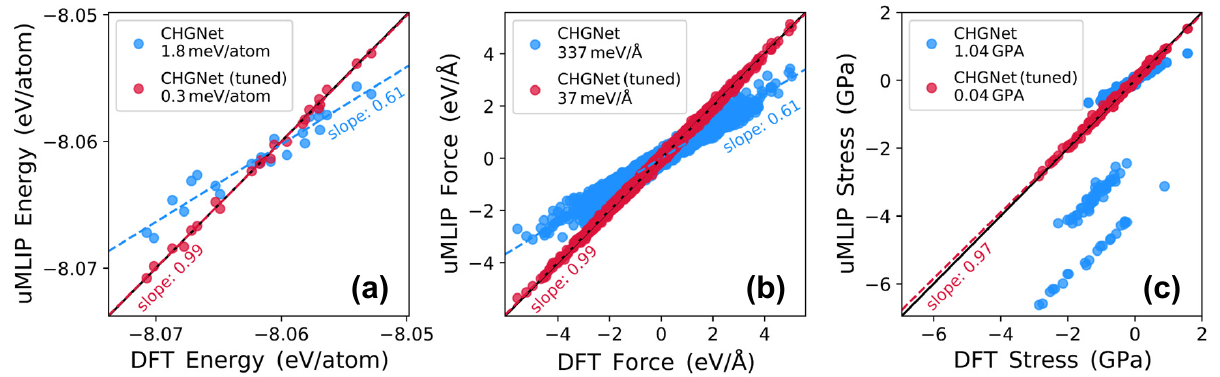}
	\caption{The parity plots of uMLIP~vs.~DFT:
		(a) energies, (b) forces, and (c) stresses.
		The respective MAEs are shown in the legends,
		and the slopes are detailed on the graph.
		The data points are derived from the TS1–-TS4 test sets (Table\ S3).
		In (a), the energies of the original and fine-tuned uMLIP
		are shifted by 86~meV/atom and 1.5~meV/atom, respectively,
		to correct for a systematic shift vs. DFT
		on an absolute scale.}
	\label{fig2}
\end{figure*}

Vanilla CHGNet yields large errors in forces and stresses of \ch{WS2} when evaluated against \emph{ab initio} data computed using the PBE exchange-correlation functional\cite{PBE_1} with D3 correction\cite{D3_correction} for vdW interactions. The mean absolute error (MAE) in force values is substantial, at 337~meV/\AA\  (evaluated on the test set with atoms randomly displaced to mimic thermal disorder).
Notably, vanilla CHGNet consistently underestimates the absolute values of forces and energies, as shown in the parity plots in Figure\ \ref{fig2}.
Here, the uMLIP energies and forces have slopes of 0.6, deviating significantly from ideal parity slopes of 1. This reflects a common problem of systematic softening of uMLIPs\cite{deng2024overcoming}, due to pre-training datasets being dominated by structures near the ground state. Predicted stresses are largely inaccurate, with MAE of 1.0~GPa, partly because CHGNet does not account for vdW interactions. When compared against PBE-level stresses (without the vdW correction), the MAE is about 0.3~GPa (we note that there is a disparity between PBE and PBE--D3 stresses, while forces and energies show perfect parity between the two functionals, see Figure S1).

Fine-tuning of CHGNet is performed using relaxation trajectories of \ch{WS2} computed at the PBE--D3 level\cite{PBE_1,D3_correction}, utilizing 7000 DFT frames. The fine-tuned CHGNet accurately reproduces the crystal structure parameters, including the vdW gap thickness, with all key values ($a$, $d_{\text{WS}}$, $z_{\text{vdW}}$, $c$) agreeing with PBE--D3 within 0.005~\AA. Moreover, it accurately reproduces PBE--D3 energies, forces, and stresses, with MAEs of 0.3~meV/atom, 37~meV/\AA, and 0.04~GPa, which is a significant improvement compared to the vanilla model (Figure\ \ref{fig2}). Fine-tuning corrects systematic softening, resulting in parity slopes close to 1. We further validated the model on MD snapshots at room and high temperatures and on strained supercells, where it maintained similar accuracy, highlighting its robustness for studying thermal disorder (see Table\ S3 in Supporting Information).

\begin{figure*}[h]
	\centering
	\includegraphics[width=0.9\textwidth]{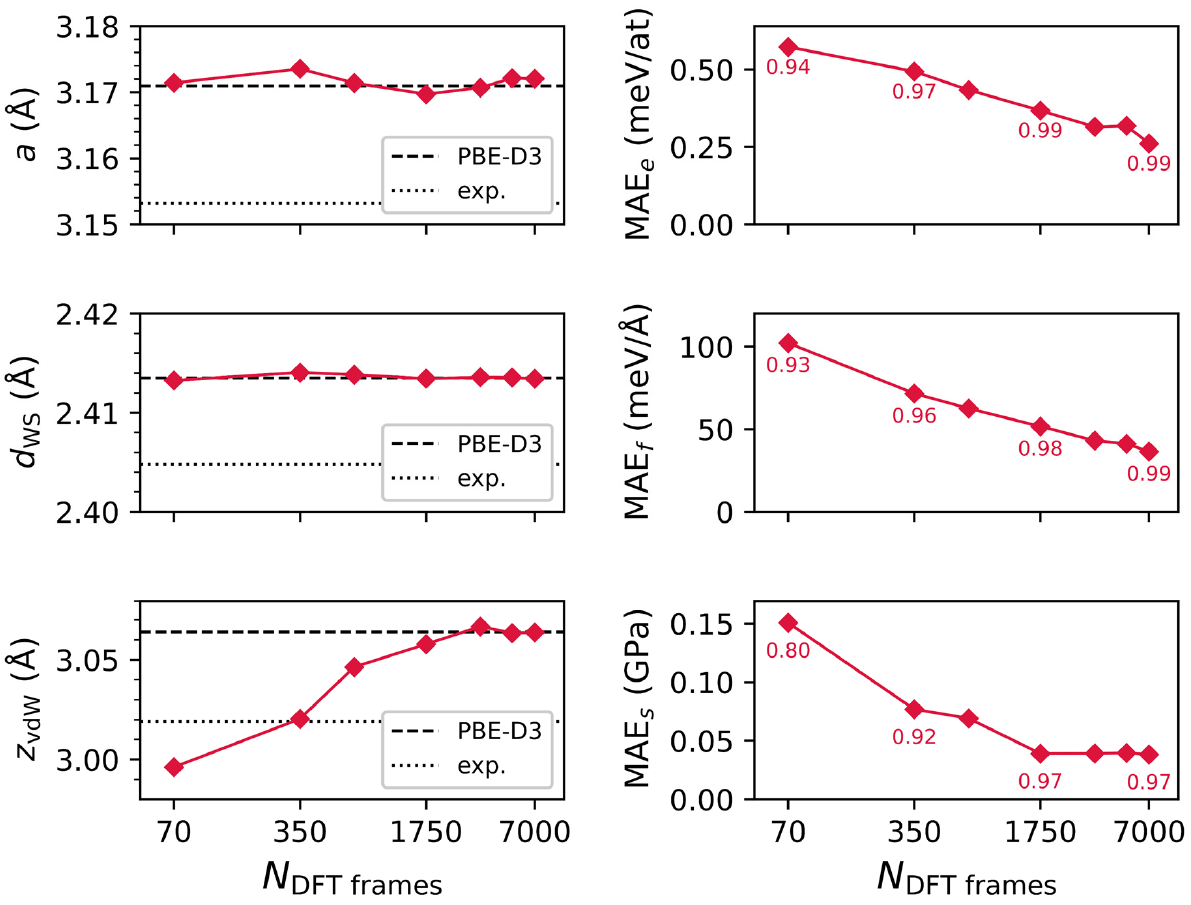}
	\caption{The accuracy of the fine-tuned CHGNet as a function of
		the number of structures in the DFT dataset
		(note, the x-axis is logarithmic).
		The left column shows the relaxed geometry parameters:
		the lattice parameter ($a$),
		the W--S bond length ($d_{\mathrm{WS}}$), 
		and the vdW gap ($z_{\mathrm{vdW}}$).
		The PBE--D3 and experimental\protect\cite{Schutte1987}
		values are indicated with
		dashed and dotted lines, respectively.				
		The right column shows MAEs for energies, forces,
		and stresses evaluated on the test sets
		TS1--TS4 (Table\ S3).
		The MAE for energies is computed after 
		shifting CHGNet energies by a constant
		to align them with DFT energies on an absolute scale.
		The numbers below the data points represent the slopes of the corresponding parity plots, where a slope of 1 would indicate perfect parity between CHGNet predictions and DFT results.
		}
	\label{fig3}
\end{figure*}

We use a relatively large dataset for fine-tuning, given that MLIPs typically need only ~500 frames to reach similar MAEs (roughly 1~meV/atom and 100~meV/\AA\ or lower)\cite{Korotaev2019thermal,Verdi2021thermal}. To test if fewer frames could achieve comparable accuracy, we evaluate progressively smaller datasets (Figure\ \ref{fig3}). Even with 70 frames, MAEs improve compared to vanilla CHGNet, reducing to 0.6~meV/atom, 102~meV/\AA, and 0.15~GPa, with softening corrected as evident from the near-ideal parity slopes of 0.94, 0.93, and 0.80 for energies, forces, and stresses. Furthermore, $a$ and $d_{\text{WS}}$ converge to PBE--D3 values within 0.01~\AA, and $z_{\text{vdW}}$ within 0.07~\AA. Including more DFT frames naturally improves performance; for example, tuning with 350 structures yields MAEs of 0.5~meV/atom, 72~meV/\AA, and 0.08~GPa, with further improvements in parity slopes.

While $a$ and $d_{\text{WS}}$ converge quickly to PBE--D3 values, it takes around 2000 DFT frames for $z_{\text{vdW}}$ to converge within 0.01~\AA\ (Figure\ \ref{fig3}). This slower convergence in $z_{\text{vdW}}$ likely stems from CHGNet not being pre-trained on vdW interactions. Explicitly incorporating D3 correction\cite{D3_correction} into the uMLIP could resolve this limitation. This result also suggests that CHGNet might feasibly learn other vdW functionals that are more complex to compute than D3.

It is valuable to consider two limiting cases: i) the number of DFT frames required to achieve ultra-high accuracy (defined here as MAEs in forces around 10~meV/\AA); and ii) the accuracy of uMLIPs trained on very small datasets, such as just a few frames. These cases represent two distinct objectives: one aiming for high accuracy, and the other focused on tuning with minimal computational resources.

Despite training on 7000 DFT frames, the MAE in forces remains at 37~meV/\AA. Previous CHGNet studies have achieved MAEs of $\sim$10~meV/\AA\ with approximately 20000 DFT frames\cite{zhong2024disorder}. Based on Figure\ \ref{fig3}, achieving MAEs of approximately 10~meV/\AA\  would require roughly a tenfold increase in DFT frames. However, we do not expand the dataset further, as additional MAE reductions have minimal impact on modeling EXAFS (discussed later), and the achieved accuracy is on par with standard \emph{ab initio} MD simulations. Indeed, while we use accurate \emph{ab initio} settings, MD simulations typically employ lower energy cut-offs or sparser $k$-point grids. Thus, a two-times sparser $k$-point grid yields MAE around 30~meV/\AA, affirming that our model attained accuracy typical of \emph{ab initio} MD.

Turning to the opposite limiting case, training on a single DFT
frame significantly improves the vanilla model,
roughly halving the MAEs in energies and forces and enhancing the
respective parity slopes,
though the improvement in stresses is marginal.
Specifically, MAEs are reduced to
0.9 $\pm$ 0.1~meV/atom,
178 $\pm$ 18~meV/\AA, and
0.7 $\pm$ 0.3~GPa, with parity slopes improving to
0.89 $\pm$ 0.05 for energies and
0.87 $\pm$ 0.04 for forces
(uncertainties estimated from using different
single DFT structures).
Therefore, fine-tuning with just a single structure leads to
a significant improvement in energies and forces compared to
the vanilla CHGNet,
which agrees with a previous study\cite{deng2024overcoming}.
Adding more structures (on the order of 10 structures)
can further enhance accuracy
(Figures\ S2--S4), 
indicating that minimal fine-tuning schemes\cite{deng2024overcoming}
may be effective for rapid refinement of uMLIPs.
However, the accuracy of uMLIPs is not necessarily a monotonous function of the number of structures (Figures\ S2--S4); using about 100 DFT frames seems to ensure robust performance.

To further strengthen our conclusions derived above, we perform a similar analysis for the isostructural 2H$_c$-MoS$_2$. The results are consistent, namely vanilla CHGNet shows pronounced softening which is remedied by fine-tuning (Figure\ S5), and about 100 structures suffice to achieve accuracy on the order of ~100 meV/\AA~ as well as to accurately reproduce the structural parameters (Figure\ S6). We also assess how the fine-tuned uMLIP reproduces related properties, such as elastic constants and phonon dispersion frequencies for both 2H$_c$-WS$_2$ and 2H$_c$-MoS$_2$ (Figures\ S7--S10). Fine-tuning on one to ten structures provides some improvement over the vanilla CHGNet; however, the deviations remain notable. For example, the $C_{11}$ elastic constant differs by up to ~100~GPa from the PBE--D3 reference values. Fine-tuning on approximately 100 structures significantly reduces these errors, which is consistent with the improvements observed in force predictions.

\subsection{Benchmarking uMLIP against EXAFS}

\begin{figure*}
	\includegraphics[width=0.9\textwidth]{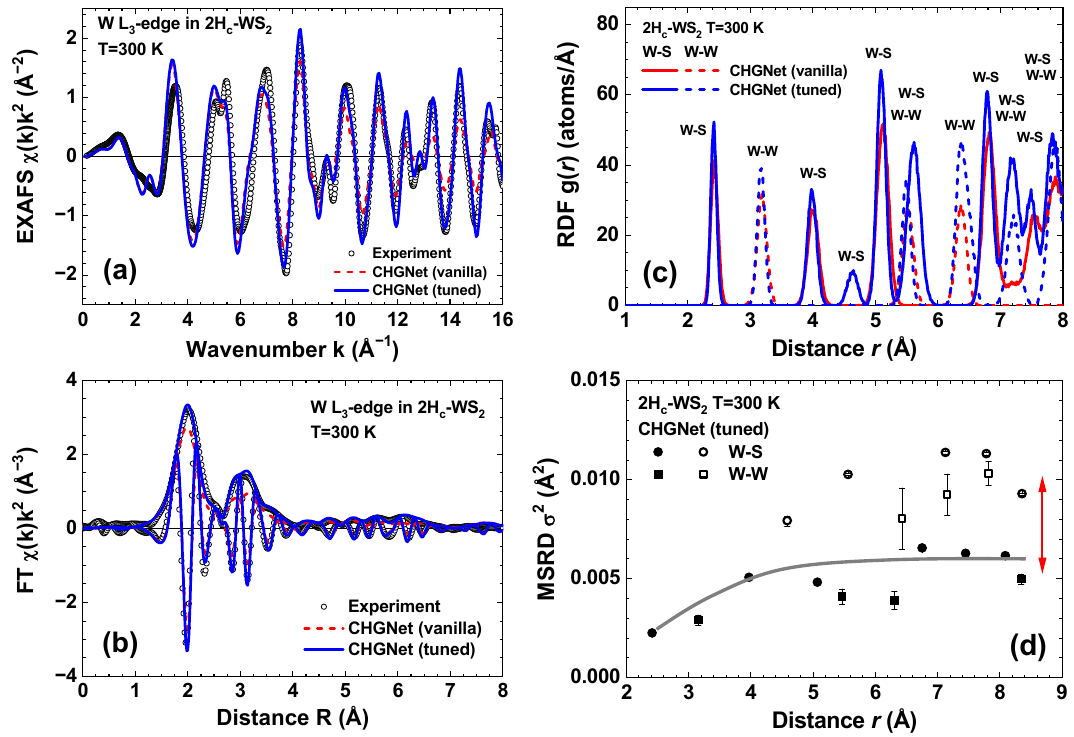}	
	\caption{Experimental (open circles) and MD--EXAFS  W L$_3$-edge EXAFS spectra $\chi(k)k^2$ (a) and their Fourier transforms (FTs) (b) for 2H$_c$-WS$_2$ at 300~K, obtained using the vanilla (dashed lines) and fine-tuned (solid lines) CHGNet potentials.	Both the moduli and imaginary parts of the FTs are shown. 
	(c)  Partial radial distribution functions (RDFs), $g_\text{W--S}(r)$\  (solid lines) and $g_\text{W--W}(r)$\ (dashed lines), obtained from MD simulations using both potentials.
   (d) 	Dependence of the mean-square relative displacements (MSRDs) on interatomic distance in 2H$_c$-WS$_2$ at 300~K according to MD simulations using the fine-tuned CHGNet  potential.    Solid symbols represent intralayer atom pairs, while open symbols represent interlayer atom pairs. The solid line serves as a guide to the eye, indicating variations in the intralayer MSRDs. Note the gap of about 0.005~\AA$^2$ between intralayer and interlayer MSRDs, highlighted by the double-headed arrow.}
	\label{fig4}
\end{figure*}

Next, we model the W L$_3$-edge spectrum using the MD--EXAFS approach,\cite{Kuzmin2009,KUZMIN2020rev} where the EXAFS spectrum is computed via \emph{ab initio} multiple-scattering theory,\cite{Rehr2000,Ankudinov1998} using MD snapshots as input geometries to account for thermal vibrations of atoms that influence the spectrum and its damping at high energies. We derive spectra from both the vanilla and fine-tuned CHGNet potentials, comparing them to the experimental EXAFS spectrum of bulk 2H$_c$-\ch{WS2} at 300~K (Figure\ \ref{fig4}). Both of the calculated spectra are rather close to each other and the experiment up to $k \approx $ 7~\AA$^{-1}$.  However, the spectrum derived from the vanilla CHGNet MD trajectory substantially overestimates damping at higher $k$-values (Figure\ \ref{fig4}(a)), suggesting increased thermal disorder\cite{Beni1976,Bohmer1979} (i.e., weaker interatomic interactions). This effect is also evident in $R$-space, in the Fourier Transform (FT) of the EXAFS spectrum (Figure\ \ref{fig4}(b)), where the vanilla CHGNet results in  substantially lower peak amplitudes around 2~\AA\ and 3~\AA\  compared to the fine-tuned CHGNet and the experiment. This aligns with the systematic softening (underestimation of forces) by vanilla CHGNet\cite{deng2024overcoming}. In contrast, the EXAFS spectrum derived from the  fine-tuned CHGNet potential shows improved agreement with the experiment in both $k$- and $R$-space.

To better understand the differences between the two potentials (vanilla vs. fine-tuned), we analyze their partial radial distribution functions (RDFs), $g_\text{W--S}(r)$ and $g_\text{W--W}(r)$, averaged over MD trajectories (Figure\ \ref{fig4}(c)). Between the two potentials, the vanilla CHGNet yields broader RDF peaks across all W coordination shells, consistent with its systematic softening. Moreover, the overestimation of the vdW gap thickness by the vanilla potential leads to significantly increased distances between atoms located in neighboring layers. Consequently, some RDF peaks calculated with the vanilla CHGNet potential appear at distances greater than 8~\AA\ and, therefore, are absent from Figure\ \ref{fig4}(c).

Analysis of EXAFS spectra provides valuable information about the pairwise correlations of atomic motion, tracked by the mean-square relative displacements (MSRDs), $\sigma^2_{ij} = \langle [(\vec{u}_i - \vec{u}_j) \cdot \hat{r}_{ij} ]^2 \rangle$, where $\vec{u}_i$ and  $\vec{u}_j$ are instantaneous displacements of atoms $i$  and $j$ from their equilibrium positions, $\hat{r}_{ij}$ is a unit vector  in the $i\text{-}j$  direction, and angular brackets denote a thermal average\cite{Beni1976}. In two-dimensional layered compounds, weak interlayer coupling  governed by vdW interactions is a key issue in both theoretical and experimental research \cite{Liu2016,Shi2018,Bian2021}. Our recent studies on 2H$_c$-MoS$_2$ \cite{Pudza2023} and 1T-Ti(V)Se$_2$ \cite{Pudza2024} demonstrate that analyzing the temperature dependence of MSRDs, derived from metal K-edge EXAFS spectra, enables the determination of interatomic force constants for both intralayer and interlayer interactions. Such information is challenging to obtain through other experimental techniques.

We obtain the MSRD values by fitting RDFs with a set of Gaussian peaks (Figure\ S11), whose areas, positions, and  variances correspond to the coordination number, interatomic distance and MSRD, respectively. The MSRDs of W--S and W--W atom pairs located within the same layer and in neighboring layers are shown in Figure\ \ref{fig4}(d) for the fine-tuned CHGNet potential. Remarkably, we observe a gap of about 0.005~\AA$^2$\ between the intralayer and interlayer MSRDs, attributed to the  weak vdW coupling between layers. Furthermore, an increase in intralayer MSRDs with increasing interatomic distance $r$ is observed due to the decreasing correlation in atomic motion of distant atoms\cite{Jeong2003,Jonane2018}. The reliability of  the obtained MSRD values is confirmed by the excellent agreement between  the  experimental and theoretical EXAFS spectra (Figure\ \ref{fig4}(a,b)). Thus, the fine-tuned CHGNet is sensitive to both intra- and interlayer interactions, predicting MSRDs consistent  with our previous studies of other layered compounds\cite{Pudza2023,Pudza2024}.

We also compare our fine-tuned CHGNet with the state-of-the-art classical potential, SWMBL-C\cite{BANDURA2023}, developed for multi-walled \ch{WS2} nanotubes and trained on various \ch{WS2} phases using advanced evolutionary optimizers\cite{Deb2013evolutionary} (Figure\ S12).  Among the two, SWMBL-C performs worse, as evident from its overestimation of experimental EXAFS damping at high $k$-values, similar to the vanilla CHGNet (Figure\ S12(a)). A comparison of the results for the SWMBL-C potential with the FTs of the experimental and fine-tuned CHGNet EXAFS spectra and the RDF for the fine-tuned CHGNet potential further indicates  that SWMBL-C overestimates the W--S bond strength and underestimates the next-nearest W--W interaction (Figure\ S12(b-d)).  While both potentials predict a gap between the intralayer and interlayer MSRDs, the value of the gap differs quantitatively (Figure\ S12(e,f)): the gap is larger for the fine-tuned CHGNet potential, which  is expected to be more accurate due to its better agreement with the experimental EXAFS spectrum.

These results demonstrate how the availability of high-quality EXAFS data (with the signal extending over a wide $k$-range) enables detailed and quantitative evaluation of interatomic potentials. The $k$-space analysis helps discern softening, while $R$-space FT analysis pinpoints which interactions are likely over- or underestimated. Instrumental to this analysis is MD--EXAFS approach\cite{Kuzmin2009,KUZMIN2020rev}, which accounts for multiple-scattering effects that  are significant across the entire $k$-range (Figure\ S13(a)) and in FT beyond 3~\AA\ (Figure\ S13(b)).

\begin{figure*}
	\includegraphics[width=0.6\textwidth]{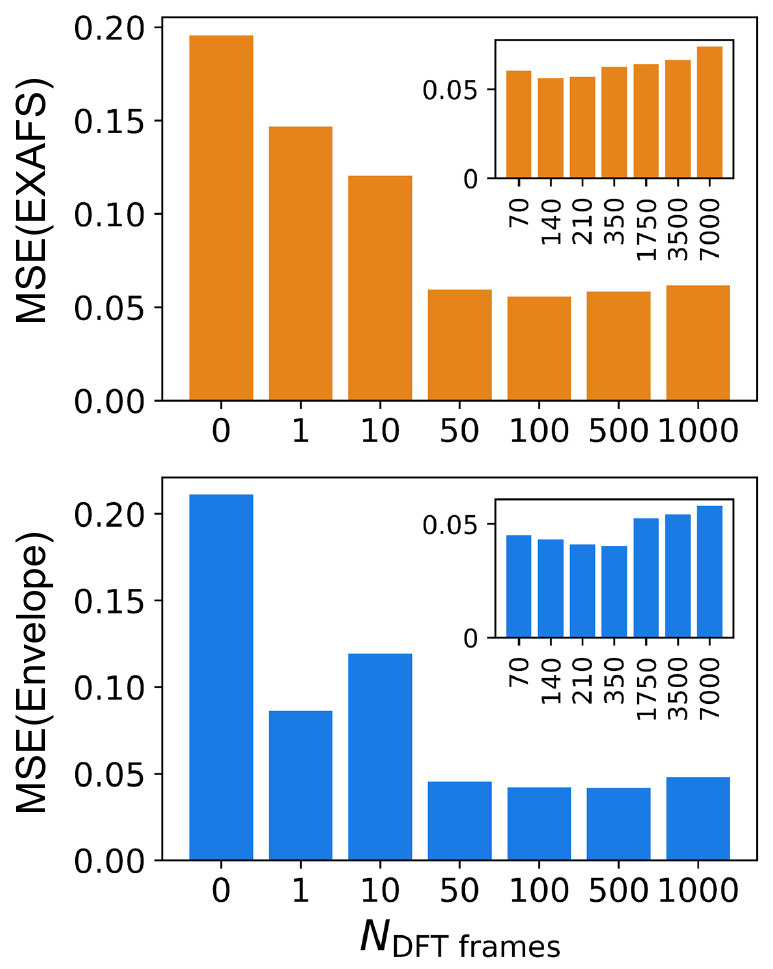}	
	\caption{The mean squared errors (MSEs) between the experimental and calculated EXAFS spectra, $\chi(k)k^2$ (upper panel), and their envelopes (lower panel), evaluated in the $k$-range of 3--16~\AA$^{-1}$. The $x$-axis represents the number of DFT frames used for CHGNet fine-tuning. The main graphs display results derived from fine-tuning on a single strain state, while the insets show results derived from seven strain states. Vanilla CHGNet corresponds to the case where $N_{\text{DFT frames}} = 0$.}
	\label{fig5}
\end{figure*}

To evaluate how the number of DFT frames impacts the accuracy of the modeled EXAFS spectrum, we use CHGNet potential fine-tuned on progressively larger DFT datasets.
The EXAFS error is calculated as the mean squared error (MSE) between the experimental and theoretical EXAFS spectra $\chi(k) k^2$ over the specified $k$-range. 
Additionally, we assess errors in the EXAFS amplitudes by calculating 
the MSE between the envelopes of the two EXAFS spectra $\chi(k) k^2$.
Both the errors between the EXAFS spectra and their envelopes are displayed in the bar chart in Figure\ \ref{fig5}. Vanilla CHGNet performs the worst, while fine-tuning on even a single DFT frame significantly reduces the errors, consistent with improvements in force accuracy. Fine-tuning on 10 frames reduces the EXAFS error further, however, the envelope error slightly increases. After fine-tuning on 50 frames, the errors stabilize, leading to no significant  changes thereafter. Interestingly, fine-tuning on a single strain state produces slightly better agreement with experimental data than using seven strain states. Increasing the number of frames leads to a negligible yet steady increase in the EXAFS error. We attribute this behavior to the slow convergence of $z_{\text{vdW}}$ toward the PBE--D3 equilibrium value, which progressively worsens agreement with the experimental structure (Figure\ \ref{fig3}). Expanding the DFT dataset improves uMLIP accuracy relative to the chosen exchange-correlation functional, but does not guarantee better agreement with experiments. Therefore, selecting a functional that accurately reproduces experimental lattice parameters is essential, with PBE--D3\cite{PBE_1,D3_correction} being the optimal choice in our case (Table\ S1).

Our tests show that approximately 100 DFT frames are sufficient for robust uMLIP performance, achieving MAE of about 100~meV/\AA\ for forces. This accuracy appears adequate for reliably reproducing the EXAFS spectrum, as further reductions in MAE  provide no significant benefit. Fine-tuning CHGNet is therefore as efficient as, or even more efficient than, training state-of-the-art MLIPs, which typically require 500 DFT frames to achieve similar accuracy.\cite{Korotaev2019thermal, Verdi2021thermal, Goodwin2024transferability} The number of frames could potentially be further reduced through efficient sampling or active learning methods (e.g., refs.\cite{Korotaev2019thermal,Zaverkin2021efficient,Podryabinkin2023mlip,qi2024stratified,Lebeda2024k_clustering}). The cost of computing 100 DFT frames is unlikely to be a bottleneck, as it is comparable to that of lengthy MD simulations. Consequently, fine-tuning is always recommended to enhance uMLIP accuracy. Notably, the fine-tuned uMLIP is able to capture vdW interactions, despite these not being explicitly represented in the original training data. This indicates that uMLIPs can effectively learn and generalize beyond their initial training domain, which is encouraging for applications involving structurally complex materials.

We also briefly evaluate other vanilla uMLIPs for their ability to predict the EXAFS spectrum. Specifically, we consider the state-of-the-art MACE uMLIP\cite{batatia2023foundation} trained on the Materials Project database (MACE-MP-0b3), as well as a variant trained on extended \emph{ab initio} datasets (MACE-MPA-0). The latter is particularly promising due to its larger training set, which includes both the Materials Project\cite{Materials_Project} and the Alexandria\cite{Alexandria_database_2024} databases, and may therefore offer superior out-of-the-box performance. The EXAFS spectra of 2H$_c$-\ch{WS2} derived from these vanilla MACE models  (Figure\  S14) show errors comparable to those of vanilla CHGNet (Figure\  S15). The addition of the D3 correction with Becke–Johnson (BJ) damping,\cite{D3_correction_with_BJ_damping} as implemented in MACE, yields a more accurate $c$ lattice parameter (Table\ S4), but does not lead to improved agreement with the experimental EXAFS spectrum. This may be due to the relatively high force MAEs, namely 256 meV/\AA\ and 160 meV/\AA\ for MACE-MP-0b3 and MACE-MPA-0, respectively. These findings further indicate that fine-tuning is important for reliable modeling of thermal disorder.

The results of the MD-EXAFS simulations at the Mo K-edge in bulk 2H$_c$-\ch{MoS2} at 300~K using the vanilla and fine-tuned CHGNet uMLIPs are shown in Figure\ S16. 
The EXAFS spectrum derived from the vanilla CHGNet MD trajectory substantially overestimates damping at higher $k$-values, leading to about 3--4 times larger  mean squared errors
(Figure\ S17) between the experimental and calculated EXAFS spectra and their envelopes than for the fine-tuned CHGNet uMLIP. Indeed, the fine-tuning improves the agreement with the experiment in both $k$- and $R$-space (Figure\ S16). Similarly to the 2H$_c$-\ch{WS2} case, about 100 DFT frames are needed to accurately reproduce the EXAFS spectrum of 2H$_c$-\ch{MoS2} (Figure\ S17).

Naturally, the number of DFT frames required for fine-tuning varies depending on the compound, the architecture of the uMLIP, and the specifics of the fine-tuning procedure. Based on our double DFT and EXAFS benchmarks, we provide a guideline for the approximate number of structures needed --- though this can likely be further optimized. In any case, we find that the required number is relatively small, which is encouraging for achieving high-level accuracy with minimal DFT input.

Finally, we note that the rigorous analysis of EXAFS spectra is a challenging task, particularly for complex materials\cite{Joress2023exafsHAE,Taheri2023SROinCCA}. Previous studies have demonstrated the significant potential of the MD--EXAFS approach\cite{Kuzmin2009,KUZMIN2020rev}, leveraging \emph{ab initio} MD\cite{Bocharov2021ZnO,Pudza2023} and MLIP-based MD\cite{shapeev2022validation} to facilitate EXAFS analysis. Given the low computational cost of DFT-based fine-tuning, we anticipate that the adoption of uMLIPs will make MD--EXAFS analysis\cite{Kuzmin2009,KUZMIN2020rev} increasingly accessible to the broader materials science community.

\section{Conclusions}

In conclusion, we have evaluated the performance of the CHGNet uMLIP\cite{deng2023chgnet} in modeling thermal disorder in bulk 2H$_{c}$-\ch{WS2} and 2H$_{c}$-\ch{MoS2}, and compared it to both \emph{ab initio} data and the experimental W L$_3$-edge and Mo K-edge EXAFS spectra at 300~K. The vanilla CHGNet systematically softens forces and energies, as evidenced by the overestimated damping of the EXAFS spectrum and the underestimated FT amplitudes. Fine-tuning the model with even a single DFT frame significantly improves accuracy in both force and energy calculations, largely mitigating this softening. The agreement with the experimental EXAFS spectrum  is also  improved, suggesting that rapid refinement of uMLIPs with minimal DFT input\cite{deng2023chgnet} is a viable approach.

Fine-tuning on $\sim$100 DFT frames provides significant performance improvements, making the CHGNet uMLIP more robust. This includes not only enhanced accuracy in forces and energies but also more precise stress calculations. The errors in forces are reduced to $\sim$100~meV/\AA, comparable to \emph{ab initio} accuracy, leading to quantitatively better agreement with the EXAFS experiment. Therefore, we recommend consistently fine-tuning uMLIPs on about 100 DFT frames, unless constrained by computational budget. Naturally, the actual number of required DFT frames depends on the compound, the uMLIP architecture, and the details of the fine-tuning procedure. In any case, this number appears to be appealingly small, whereas state-of-the-art MLIPs  typically require about 500 DFT frames to achieve a similar level of accuracy\cite{Korotaev2019thermal, Verdi2021thermal, Goodwin2024transferability}.

Finally, given the low computational cost of uMLIP fine-tuning, we anticipate that their widespread adoption will facilitate the analysis of EXAFS spectra through the MD--EXAFS approach,\cite{Kuzmin2009,KUZMIN2020rev} making it increasingly accessible to the broader materials science community. Future studies should further evaluate the performance of uMLIPs in modeling the structure and thermal disorder of more complex materials, such as those with low symmetry, significant lattice distortions, or compositional complexity.

\begin{acknowledgement}

P.\v{Z}. acknowledges the support of the project
No. 1.1.1.9/LZP/1/24/016 by European Regional Development Fund.
I.P. thanks the support of the Latvian Council of Science project No. LZP-2023/1-0528.
The experiment at the PETRA III synchrotron was performed within proposal No. I-20170739 EC.
We would like to thank Dr. Edmund Welter for his assistance in using the P65 beamline. The synchrotron experiment has been supported by the project CALIPSOplus under the Grant Agreement 730872 from the EU Framework Programme for Research and Innovation HORIZON 2020.
	
\end{acknowledgement}

\begin{suppinfo}
The Supporting Information is available free of charge.
	
	\begin{itemize}
		\item Justification for the exchange-correlation functional choice; crystallographic data; additional uMLIP tests; extraction of MSRDs; comparison with the SWMBL-C force field; contribution of multiple-scattering effects.
		\item TS1--TS4 test sets (structures, energies, forces, stresses) for uMLIP benchmarking are available at https://zenodo.org/records/16420574
	\end{itemize}
	
\end{suppinfo}

\clearpage
\newpage

\clearpage
\newpage
\bibliography{xafs,ws2,vasp_etc,mlip}

\end{document}